\documentclass[runningheads]{llncs}
\usepackage{graphicx}
\usepackage[ruled,vlined]{algorithm2e}
\begin{document}
\title{Blind stain separation using model-aware generative learning and its applications on fluorescence microscopy images }
%
%


\author{Xingyu Li}
\institute{University of Alberta, Edmonton, Alberta, Canada \\ 
\email{xingyu@ualberta.ca}}

%
\maketitle              
\begin{abstract}
Multiple stains are usually used to highlight biological substances in biomedical image analysis. To decompose multiple stains for co-localization quantification, blind source separation is usually performed. Prior model-based stain separation methods usually rely on stains' spatial distributions over an image and may fail to solve the co-localization problem. With the advantage of machine learning, deep generative models are used for this purpose. Since prior knowledge of imaging models is ignored in purely data-driven solutions, these methods may be sub-optimal. In this study, a novel learning-based blind source separation framework is proposed, where the physical model of biomedical imaging is incorporated to regularize the learning process. The introduced model-relevant adversarial loss couples all generators in the framework and limits the capacities of the generative models. Further more, a training algorithm is innovated for the proposed framework to avoid inter-generator confusion during learning. This paper particularly takes fluorescence unmixing in fluorescence microscopy images as an application example of the proposed framework. Qualitative and quantitative experimentation on a public fluorescence microscopy image set demonstrates the superiority of the proposed method over both prior model-based approaches and learning-based methods. 

\end{abstract}
\section{Introduction}
In biomedical image analysis, multiple stains are usually used to highlight  biological substances of interest and their interactions in tissue samples for quantitative and qualitative analysis. Fluorescence staining and histopathology staining are the representative staining protocols widely adopted in tissue sample analysis. One issue in stained images is mixing/blurred colors due to co-localization of stained biological substances. To regain the information provided by the contrast of individual stains, stain separation that facilitates the quantitative evaluation of degree of co-localization is highly desired in biological research.

Stain unmixing, or stain separation, is a specific source separation process that separates the mixing stains in a biomedical image into a collection of single-stained planes/images. Model-based unmixing algorithms rely on specific mixing models, which can be formulated as either linear or nonlinear. For instance, fluorescence imaging follows a linear mixing model \cite{ref_FI} and H\&E stained histopathology imaging is characterized by the non-linear Beer–Lambert law \cite{ref_bll}. In early literature, stain unmixing is usually formulated as a model-based inverse problem and many classic matrix decomposition techniques, for instance, ICA \cite{ref_ICA} and SVD \cite{ref_SVD}, were deployed. Later, to address such a underdetermined source separation problem, various regularization terms such as sparsity \cite{ref_sparsity} and non-negative constraint \cite{ref_NMF}\cite{ref_NMF2} are introduced in model-based methods. However, these statistics approaches heavily rely on stains' distributions in an image and usually fail for weak stained instances. 

With the advance of deep learning, blind source separation (BSS) problems are now tackled by generative models \cite{ref_speechGAN}\cite{ref_imageGAN}. Particularly for stain separation, the special case of BSS, deep models (such as U-Net \cite{ref_Unet} and generative adversary networks (GAN) \cite{ref_GAN}\cite{ref_pix2pix}) can be built based on a set of training samples that consists of pairs of multi-color biomedical images and their decompositions. Different from the model-based methods relying on the statistics of stains in an image, deep generative models learn the decomposition function from the training set and thus are less sensitive to stains' distribution in a specific image.

In this work, a novel end-to-end model-aware GAN-based BSS framework is proposed for stain separation in biomedical images. Unlike the previous deep learning BSS works which are only data-driven, the specific imaging model is incorporated in the proposed scheme. Briefly, in the proposed framework, multiple deep nets are trained to generate single-stained images. By exploiting the imaging model, a model-relevant adversarial loss is used to couple individual generator nets in the learning process. Additionally, a new training algorithm is innovated, where the model-relevant adversarial loss is involved in training and a learning strategy is investigate to avoid inter-generator error propagation. Experimentation on fluorescence separation in biomedical images demonstrates that prior knowledge of imaging model improves the learning performance and outperforms prior arts. In summary, the contribution of the paper are as follows:
\begin{itemize}
	\item To our best of knowledge, it is the first time in literature to innovate a model-aware deep learning framework for biomedical image stain separation. 
	\item We incorporate generative learning and physical imaging models within the proposed framework. The solution is robust to variations in stains' statistics in an image. Compared to prior deep models, the proposed method generates decomposition results in higher quality.
	\item A novel end-to-end training algorithm is proposed to optimize the learning efficiency. The model-relevant loss is incorporated in training to couple individual generators toward the optimal solution and accelerate the learning process in the late training phase.
\end{itemize}

\section{Methodology}
\textbf{Framework Overview: }Stain separation decomposes a multi-stained image $I$ into multiple images $I_i$, each containing biological substances of interest stained by one stain. Fig. 1 depicts the overview system architecture. In the framework, $N$ generator nets are used for single-stained image generation. After the image syntheses module which takes the corresponding physical imaging model as the basis, a discriminator net is incorporated to yield a model-relevant adversarial loss. Note that the generator $G_i: I\rightarrow I_i$ and discriminator $D$ are trainable and that the synthesis module $S:\{I_1,...,I_N\}\rightarrow I$ adopts a specific imaging model of the on-hand problem to couple the $N$ generators. 

\begin{figure}
	\includegraphics[width=\textwidth]{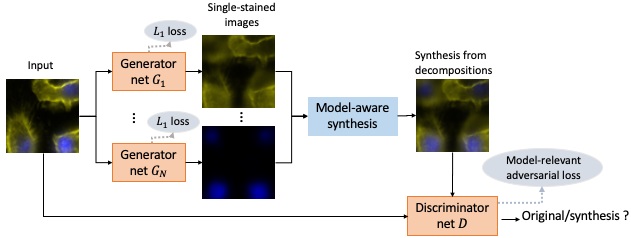}
	\caption{Overview of the model-aware stain separation learning architecture, where the orange modules represent learning models with trainable parameters and the blue module is defined by the deterministic imaging model. During training, each $G_i$ first targets to minimize the $L_1$ loss between the ground-truth and its generated single-stained image only; when $L_1$ losses are small, the model-relevant adversarial loss is incorporated to accelerate generators' optimization. During test, $G_i$ are used for stain separation.} \label{fig1}
\end{figure}

It is noteworthy that though the proposed framework adopts the adversarial loss, it is distinct from the classical GAN scheme. Instead of using multiple GAN nets to generate individual single-stained images, only one discriminator $D$ is used to couple all generators in the learning processing. The model-aware adversarial loss drives all generators to work hard together to compete against the discriminator. Consider the subsystem between the input $I$ and its synthesis $\hat{I}$, let's denote it as $F\circ S$ where $F=\{G_1,...,G_N\}: I\rightarrow\{I_1,...,I_N\}$ is the decomposition function contributed by all generator nets and $\circ$ represents module concatenation. Then the proposed model-aware scheme can be rephrased as to find the decomposition function $F$ such that $F\circ S$ is an identity function, i.e. $F\circ S: I\rightarrow I$. To this end, the proposed model-aware adversarial loss is designed as a function of $F\circ S$, uniting all generators in training.
 
One may notice that error associated with $G_i$ may propagate to other generators $G_{j}$ for $j\neq i$ through the model-aware adversarial loss, consequently causing failure. To avoid such inter-generator confusion in learning, a novel end-to-end training algorithm is proposed to decouple generators at the early training phase. \\

\noindent\textbf{Loss Functions:}
As illustrated in Fig. 1, there are two categories of loss functions in the proposed scheme. For each generator $G_i$, an $L_1$ loss is used to quantify the difference between the generated single-stained image $\hat{I}_i$ and its ground truth $I_i$, i.e.
\begin{equation}
 \mathcal{L}_{L_1}^{i} (G_i)=\mathbf{E}[|I_i-\hat(I_i)|_1]=\mathbf{E}[|I_i-G_i(I)|_1].
\end{equation}
The reason that $L_1$ loss is preferred over the common $L_2$ loss is that $L_2$ loss tends to generate smooth/blue images. 

For the proposed architecture, we innovate a model-aware adversarial loss that couples all $G_i$s. Based on the design, $F\circ S: I\rightarrow I$ is equivalent to an autoencoder with specific intermediate representations defined by the decomposition ground truth. Then $F\circ S$ and $D$ compose a conditional GAN, where $F\circ S$ targets to fool the discriminator $D$ and $D$ tries to improve its classification performance. Hence, the model-aware adversarial loss is
\begin{equation}
 \mathcal{L}_{cGAN} (D, F\circ S)=\mathbf{E}[\log D(I,I)]+\mathbf{E}[\log(1-D(I,F\circ S(I)))]. \end{equation}
Note that since the synthesis module $S$ is deterministic and well-defined by a specific imaging model, all trainable parameters in the subsystem $F\circ S$ originate from $F=\{G_1,...,G_N\}$. This suggests that each generator $G_i$ ties to minimize the adversarial loss in Eqn. (3) while the discriminator D tries to maximize it.

The advantage of the model-aware adversarial loss is to couple all generators for learning augmentation. But it also brings in a risk that the inter-generator confusion collapses the whole BSS system. Specifically, stain decomposition is an under-determined problem as there may be numerous sets of $\{\hat{I}_i,...,\hat{I}_N\}$ satisfying  the imaging model $S$. Hence, if the single-stained image $\hat{I}_i$ generated by $G_i$ greatly deviates from its true value $I_i$, the corresponding error may propagate to other generators through the adversarial loss. To address this issue, rather than directly training $F\circ S$ towards the identity function via the adversarial loss, we enforce the generators $G_i$ focusing on estimation of decomposition ground-truth by $L_1$ loss in Eqn. (2); $G_i$s are coupled through the model-aware adversarial loss only after generators have converged to the region that is close to the ground truth $I_i$. In this way, the learning process will be accelerated even at the late phase of training. In sum, the overall loss function is 
\begin{equation}
\mathcal{L}(D, F\circ S) =\sum_{i=1}^N\mathcal{L}_{L_1}^i+\lambda \mathcal{L}_{cGAN} (D, F\circ S),
\end{equation}
where $\lambda$ is a hyper-parameter to weight the adversarial loss in the overall target function. In the early training phase, $\lambda=0$ to enable the generators to learn from the decomposition ground truth $I_i$. In the late phase of training when $L_1$ loss is small, $\lambda > 0$ to couple all generators for parameter optimization. Therefore, the optimal solution to the problem is 
\begin{eqnarray*}
F^* =  \arg \min_F \max_D \mathcal{L} (D, F\circ S) 
= \arg \min_F \max_D \left[\sum_{i=1}^N\mathcal{L}_{L_1}^i+\lambda \mathcal{L}_{cGAN} (D, F\circ S)\right].
\end{eqnarray*}

\textbf{Training Algorithm:} In this work, the standard procedures in GAN update \cite{ref_GAN} is followed - we alternate between one gradient descent step on discriminator $D$ and then one step on generators $G_i$. The specific end-to-end training procedure is described in Algorithm 1.

\begin{algorithm}[H]
 	\SetAlgoLined
 	\KwIn{Training data set: multi-color images $I$s and its BBS groud truth $I_i, i=1,...,N$, the number of sourse $N$, adversarial loss weight $\lambda$, and adversarial loss involvement parameter $\alpha$}
 	\KwOut{Parameters of generators $\theta_{G_i}$ and deccriminator $\theta_D$ }
 	\For{number of training iterations}{
 		Sample minibatch of $m$ training data $\{I^{1},...,I^{m}\}$ \;
 		Update the discriminator $D$ by ascending the stochastic gradient:
 		$\nabla_{\theta_D} \frac{1}{m}\sum_{j=1}^m \left[\log D(I^j,I^j)+\log(1-D(I^j,F\circ S(I^j))) \right]$ \;
 		Sample minibatch of $m$ training data $\{I^{1},...,I^{m}\}$ and $\{I_{i}^{1},...,I_{i}^{m}\}$ \;
 		\eIf{first $100-\alpha$ percent iterations}{
 			Update the generators $G_i$ by ascending the stochastic gradients:
 			$\nabla_{\theta_{G_i}} \frac{1}{m}\sum_{j=1}^m |I_i^j-G_i(I_i^j)|_1$ \;
 		}{
 			Update the generators $G_i$ by ascending the stochastic gradients:
 			$\nabla_{\theta_{G_i}} \frac{1}{m}\sum_{j=1}^m \left[|I_i^j-G_i(I_i^j)|_1 + \lambda \log(1-D(I^j,F\circ S(I^j))) \right]$ \;
 		}
 	} 	
 \caption{Minibatch stochastic gradient descent training of the proposed model-aware BBS learning}
\end{algorithm}


\section{Fluorescence Unmixing}
Fluorescence microscopy is a technique used to visualize and identify the distribution of proteins or other molecules of interest stained with fluorescent stains. Because fluorescence microscopy is able to precisely distinguish individual antigens, it is the most popular cell and molecular biological research tool to study dynamical changes in the cell. To highlight multiple molecules of interest within a sample and analyze their interactions, multi-color fluorescence microscopy is used. 

Fluorescence unmixing aims to decompose a multi-color fluorescence microscopy image into multiple images, each containing proteins or other molecules of interest stained with one fluorescent stain only. Mathematically, given that there are $N$ types of fluorescent stains in an image and let $V_i$ and $D_i$ represent the $i^{th}$ fluorescent stain's spectrum vector and staining density map that consists of stain proportions over all pixels, a fluorescence microscopy image can be formulated as a linear combination of the $N$ staining density maps \cite{ref_FI}:
\begin{equation}
I=V\times D=[V_1,...,V_N]\times [D_1,...,D_N]^T = \sum_{i=1}^N V_iD_i=\sum_{i=1}^{N}I_i,
\end{equation}
where $I$ and $I_i$ are the observed multi-color fluorescence microscopy image and its $i^{th}$ decomposed instance, i.e. the single-stained image associated with the $i^{th}$ fluorescent stain, respectively. $[.]^T$ represents matrix transpose.

To achieve fluorescent stain separation, model-based approaches in literature exploit the imaging model in Eqn. (4) and estimate stains' spectrum matrix $V$ from a query image. Then staining density maps $D$ is obtained by matrix inverse operation, and single-stained image $I_i$ is derived by $I_i=V_iD_i$. Among matrix inverse methods, NMF based approaches achieves top performance \cite{ref_NMF}, where non-negative properties of $D$ and $V$ physically are considered during matrix factorization. However, NMF-based solutions usually have weak ability to handle molecules' co-localization. 

In this paper, we use fluorescence separation as the application of the proposed framework. Different from model-based approaches that searches optimal $V$ and $D$ first, the proposed method takes Eqn. (4) as the image synthesis model and directly generates a set of seperation results $\{I_i\}$. In the next section, we evaluate the proposed stain separation framework on fluorescence images and compare its performance with prior arts.

\section{Experimental Evaluation}
\subsection{Experimental Setup}
\textbf{Dataset:} We use image set BBBC020 from the Broad Bioimage Benchmark Collection \cite{ref_BBBC} in this study. In the high-resolution RGB-format fluorescence microscopy images of murine bone-marrow derived macrophages from C57BL/6 mice, nuclei were labeled by DAPI in blue and cell surface was stained by CD11b/APC in yellow \cite{ref_dataset}. In addition to multicolor images containing two fluorescent stains, single-stained images (i.e. DAPI images and CD22b/APC images) are provided in the image set. Since the cells possess an irregular shape and some of the macrophages touch or overlap each other, this dataset can be used to assess an algorithm's ability to deal with co-localization cell data \cite{ref_BBBC}. In this study, each high-resolution image is divided into 30 $256\times256$ patches. In this way, we collect 2250 image patches coming from 750 decomposition patch pairs, each pair containing one multi-color image and its source separation ground truth (i.e. one DAPI image and one CD11b/APC image). Then we randomly picked 10\% image pairs as test cases and the rest are used to train the proposed BBS model.

\noindent\textbf{Network Architectures \& Hyperparameter Setting:} In this study, we adopt the generators and discriminator from those in \cite{ref_pix2pix}. Specifically, U-Net256 \cite{ref_Unet256} is used to realize the generators, where the input tensor has a size of $256\times 256\time 3$. Compared to the encoder-decoder structure, the skip path in U-Net helps to maintain cell/molecule structures in images. Regarding to the discriminator, a $70\times70$ PatchGAN which aim to classify whether $70\times70$ overlapping image patches are real or fake is adopted. It is noteworthy that the U-Net and the PatchGAN are applicable on any color images whose resolution is larger than $256\times 256$. To train the model, minibatch SGD realized by Adam \cite{ref_adam} is applied, with a learning rate of 0.0002, and momentum parameters $\beta_1 = 0.5, \beta_2 = 0.999$. For hyperparameters relevant to model-aware adversarial loss, $\alpha = 75, \lambda = 0.01$.

\noindent\textbf{Comparison with State-of-the-Art Methods:} NMF based methods achieved top performance in model-based fluorescence microscopy image separation. Hence, we include the NMF based method \cite{ref_NMF} in this comparison evaluation and use it as the baseline. Since there is no existing learning based approaches proposed for biomedical image stain separation, two common generative learning models are evaluated. In specific, since we have two fluorescence stains in the image set, either 2 U-Nets \cite{ref_Unet256} with the $L_1$ loss or 2 pix2pix GAN \cite{ref_pix2pix} are used.

\noindent\textbf{Evaluation Metrics:} Both qualitative and quantitative evaluations are conducted. In qualitative comparison, spectral separation results generated by different methods are visually examined. We also compute three major image quality metrics (MSE, PSNR, and SSIM \cite{ref_ssim}) between the decomposition results and the ground truth for quantitative comparison.

\subsection{Results and Discussions}
\begin{figure}
	\includegraphics[width=\textwidth]{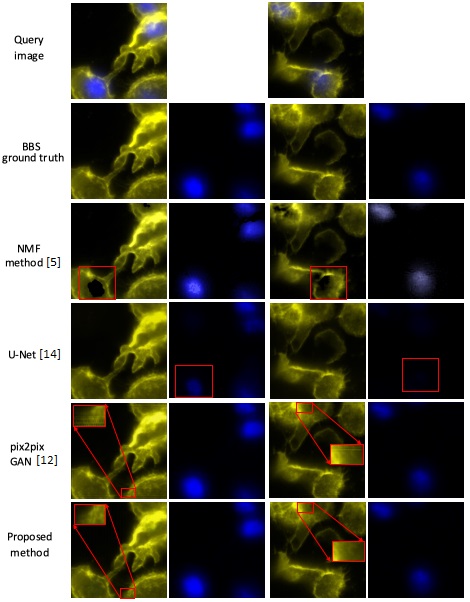}
	\caption{Samples of fluorescence separation using different methods. Images in the second row correspond to BSS ground truth. The NMF method \cite{ref_NMF} and U-Net [14] may fail to handle co-localization; Images generated by the pix2pix GAN [12] appears with noticeable check-block effects, while the proposed method generates smooth results as shown in the figure (It is recommended to compare the results between pix2pix GAN and the proposed method by enlarging the figure in PDF).} \label{fig2}
\end{figure}

Fig. 2 presents examples of stain separation obtained by different methods. As illustrated in the figure, the NMF-based method \cite{ref_NMF} and U-nets \cite{ref_Unet256} fail to separate co-localized fluorescent stains in the images. Though the pix2pix GAN \cite{ref_pix2pix} obtains better results, block-check artifacts which usually occur in GAN models are observed, especially along image boundaries. Compared to prior arts, the proposed method yields the top source separation results.

Table 1 records the quantitative evaluation of the examined methods. For the first column which corresponds to MSE, smaller values suggest better decomposition results. For PSNR and SSIM, the large the values are, the better the performance is. From the table, all three metrics advocates to the superiority of the proposed method, and the pix2pix GAN ranks the second despite of its check-block artifact. This quantitative comparison is consistent with our qualitative visual examination.

\begin{table}
	\caption{Quantitative evaluation of BBS via major image quality metrics, where the best values are marked black.}\label{tab1}
	\centering
	\begin{tabular}{|p{1.5cm}|p{2.5cm}|p{2.5cm}|p{2.5cm}|p{2.5cm}|}
		\hline
		&  NMF method \cite{ref_NMF} & U-nets \cite{ref_Unet256} & pix2pix GAN \cite{ref_pix2pix} & proposed method\\
		\hline
		MSE &  73.65 & 88.13 & 8.86 & \textbf{5.85} \\ \hline
		PSNR &  30.67 & 32.97 & 39.13 & \textbf{41.10} \\ \hline
		SSIM & 0.92 & 0.89 & 0.95 & \textbf{0.97} \\
		\hline
	\end{tabular}
\end{table}

The good performance of the proposed method is due to two reasons. First, let's compare the U-Net based method and the proposed method. It should be noted that both approaches adopt the U-Net as the generators. The major difference is that the proposed method introduces a model-aware adversarial loss in training. Inherent from GAN, the discriminator always tries to find a better objective function that distinguishes the reconstruction and the original input. As a result, instead of passively approaching the decomposition ground truth by $L_1$ loss, the generator works hard to compete against the discriminator. Second, let's focus on the pix2pix GAN \cite{ref_pix2pix} and the proposed method. Since both approaches adopt the adversarial game in training, they outperform the U-Net based model. But different from the model that needs 2 independent pix2pix GAN to generate the spectral decomposition results, the proposed method innovates the use of one discriminator to couple all generators via the fluorescence microscopy images. Because of the prior knowledge uniting all generators in the scheme, the proposed method has less freedom in image generation compared to the pix2pix GAN method.  

\section{Conclusions} 
This study proposed a novel deep learning based framework for BSS in biomedical images. In particular, it utilized the physical imaging model to augment the generative learning. In addition, the learning scheme was further empowered and regularized by a model-aware loss. Both qualitative and quantitative experiments had validated the efficiency of the proposed method by public fluorescence microscopy image set. Future work would aim at evaluating the proposed stain unmixing scheme on other BSS scenarios such as multi-stained histo-pathology images.

%
%
%
%

\end{document}